\documentclass[a4paper,fleqn]{cas-sc}

\usepackage[compress,numbers]{natbib}
\usepackage{footmisc}
\usepackage{subfigure}
\usepackage{multirow}
\usepackage{graphicx}
\usepackage{breakurl}

\def\tsc#1{\csdef{#1}{\textsc{\lowercase{#1}}\xspace}}
\tsc{WGM}
\tsc{QE}

\begin{document}
\let\WriteBookmarks\relax
\def\floatpagepagefraction{1}
\def\textpagefraction{.001}



\title [mode = title]{Measurement of reactor thermal neutron fluence of NTD-Ge by activation High-Purity Ge itself}  

\tnotemark[1] 

\tnotetext[1]{Supported by the National Natural Science Foundation of China(12005225, 11721505, 11625523, 61571411). the Fundamental Research Funds for the Central Universities (WK2030000024), the State Key Laboratory of Particle Detection and Electronics(SKLPDE-ZZ-202012)
} 

%

\author[1,2]{Zhao Kangkang}



\affiliation[1]{organization={Department of Modern Physics, University of Science and Technology of China},
            addressline={}, 
            city={Hefei},
            postcode={230026}, 
            state={Anhui},
            country={China}}

\author[1,2]{Xue Mingxuan}
\cormark[1]
\ead{xmx@ustc.edu.cn}
\cortext[1]{Corresponding author}
\affiliation[2]{organization={State Key Laboratory of Particle Detection and Electronics},
            addressline={}, 
            city={Hefei},
            postcode={230026}, 
            state={Anhui},
            country={China}}

\author[1,2]{Peng Haiping}
\cormark[1]
\ead{penghp@ustc.edu.cn}



\begin{abstract}
In neutron transmutation doped germanium, the thermal neutron fluence of reactor irradiation is as high as 10$^{18}$~n$\cdot$cm$^{-2}$. For radiological safety reasons, general Co or Au neutron flux monitors cannot be easily used. We have experimentally demonstrated the feasibility of measuring the X-rays emitted by the NTD-Ge itself to determine the absolute thermal neutron fluence for the first time. A Micro-Megas Detector (MMD) and a Silicon Drift Detector (SDD) are set up to detect the tagging KX-rays with 9.2 keV and 10.3 keV cascading from the decays of $^{71}$Ge. Combined the detection efficiencies calculated by GEANT4, neutron fluence results given with proper accuracy by MMD and SDD are in a good agreement with each other. 
\end{abstract}

\maketitle
\section{Introduction}
There are many experimental researches of reactor neutron irradiation in structure material, neutron therapy and particularly in semiconductor devices\cite{barabash2000neutron,hasegawa2013neutron,bortolussi2007thermal,janus1976application,shlimak1999neutron}. Neutron Transmutation Doped (NTD) technique can form a uniform doping in material via thermal neutron capture and subsequent radioactive decay, is a important application of neutron irradiation\cite{shlimak1999neutron,larrabee2013neutron}. By controlling the thermal neutron fluence, NTD semiconductor with desired dopant concentration can be massive reproduced. The thermal neutron fluence at the point of the sample located is crucial to the final performance of NTD devices, and requires accurate measurement.

A general method to determine the thermal neutron fluence in experiments is based on gamma-spectrometry of an activated nuclide of the monitor foil (called as the extra monitor) located near the sample being irradiated\cite{kohler1971determination,simonits1976zirconium,de1991calibration}. Cobalt foil with a known quantity of nuclide $_{27}^{59}$Co is most widely used as extra monitor\cite{kohler1971determination}, which is free of interfering reactions due to the 100 percent isotopic abundance of $^{59}$Co, and $^{59}$Co has a precise nuclear parameter. An undesired point is that $^{60}$Co is a long-lived radioactive nuclide, which is environmentally difficult to deal with. In fact, people usually could find out a proper isotope $_{Z}^{A}X$ in the irradiated sample itself, which has proper thermal neutron cross section of $_{Z}^{A}X$($n,\gamma$)$_{Z}^{A+1}Y$ , and proper half\-life time of nuclide $_{Z}^{A+1}X$, also proper rays ($\gamma$ or X) which are tagging the nuclide $_{Z}^{A+1}X$. The sample itself can be one of the best monitors (call as Self-monitor) of neutron fluence which is a real one free from the extra monitor’s disturbing. Taking the NTD Germanium (NTD-Ge) as an example, $^{70}$Ge (see the first line of the TABLE~\ref{tab:table0905_1}) of Ge sample could be chosen as a qualified monitor’s nuclide, Its thermal neutron capture cross section is 3.05 b, much smaller than one of $^{59}$Co (37 b)\cite{201889}. The validity of the formula~\ref{eq:0905_1} ($\sigma\varphi\ll \lambda$) would limit the thermal neutron flux to be monitored 1 order higher using $^{70}$Ge than $^{59}$Co; In addition, X-ray of $^{71}$Ga for tagging $^{71}$Ge is safer to be handled for body comparing with MeV’s $\gamma$ of $^{60}$Co; Finally the ways to deal with radioactive waste for $^{71}$Ge ($T_{\frac{1}{2}}$=11.43 d) are more convenient than one for $^{60}$Co ($T_{\frac{1}{2}}$=5.27 y).   
In this work, we firstly using high-purity germanium (HP-Ge) itself as monitor for thermal neutron irradiated Ge by measuring the X-rays accompanying to the $^{71}$Ge decays. The irradiated Ge sample itself has strong absorption for the low-energy X-rays, and the GEANT4 Monte Carlo simulation tool can accurately give the contribution of the interaction.
\newpage
\section{Principles}\label{sec::0905_2}
\begin{table*}[h!]
\caption{\label{tab:table0905_1}The natural isotopic abundance of Ge, the thermal neutron capture cross section for Ge isotopes and the radioactive decay process to the dopant isotopes.}
\begin{tabular*}{\tblwidth}{@{}LLLLL@{}}
\toprule
\multirow{2}{*}{Natural isotopes}&Abundance\cite{isotopicdata}&\multicolumn{1}{c}{\multirow{2}{*}{Doped process}}&Neutron capture&\multirow{2}{*}{Type}\\
&(\%)&&cross-section\cite{201889} (b)&\\
\midrule
$^{70}_{32}{\rm Ge}$&20.57&$^{70}_{32}{\rm Ge}(n, \gamma)\rightarrow\ ^{71}_{32}{\rm Ge}\xrightarrow{\text{EC, }T_{1/2}=11.43 \text{d}}\ ^{71}_{31}{\rm Ga}$&3.05(13)&Acceptors, p-type\\
$^{72}_{32}{\rm Ge}$&27.45&$^{72}_{32}{\rm Ge}(n, \gamma)\rightarrow\ ^{73}_{32}{\rm Ge} (\text{stable})$&0.89(8)&-\\
$^{73}_{32}{\rm Ge}$&7.75&$^{73}_{32}{\rm Ge}(n, \gamma)\rightarrow\ ^{74}_{32}{\rm Ge} (\text{stable})$&14.7(4)&-\\
$^{74}_{32}{\rm Ge}$&36.50&$^{74}_{32}{\rm Ge}(n, \gamma)\rightarrow\ ^{75}_{32}{\rm Ge}\xrightarrow{\beta^{-}, T_{1/2}=82.8 \text{min}}\ ^{75}_{33}{\rm As}$&0.36(4)&Donors, n-type\\
\multirow{2}{*}{$^{76}_{32}{\rm Ge}$}&\multirow{2}{*}{7.73}&$^{76}_{32}{\rm Ge}(n, \gamma)\rightarrow\ ^{77}_{32}{\rm Ge}\xrightarrow{\beta^{-}, T_{1/2}=11.3 \text{h}}\ ^{77}_{33}{\rm As}$&\multirow{2}{*}{0.055(2)}&\multirow{2}{*}{Donors, n-type}\\
&&\multicolumn{1}{r}{$\xrightarrow{\beta^{-}, T_{1/2}=38.8 \text{h}}\ ^{77}_{34}{\rm Se}$}\\
\bottomrule
\end{tabular*}
\end{table*}
The processes associated with NTD-Ge are shown in TABLE~\ref{tab:table0905_1}, three kinds of radionuclides $^{71}$Ge, $^{75}$Ge, $^{77}$Ge are produced. Both $^{75}$Ge and $^{77}$Ge could be tagged by their characteristic cascading $\gamma$-rays. But $^{75}$Ge nuclide could not be as a monitor nuclide because they have been almost depleted after about 2 days ($\simeq 35\cdot T_{1/2}(^{75}\text{Ge})$) moved out from the reactor, when the radioactivity of the sample decays down to the level in which people are allowed to work with. $^{77}$Ge nuclide could be a proper monitor nuclide when people detect the characteristic $\gamma$-rays of 264~keV within 1 week ($\simeq 15\cdot T_{1/2}(^{77}\text{Ge})$) to tag the activities of $^{77}$Ge. In our views, the reaction channel $^{70}{\rm Ge}(n, \gamma)\rightarrow\ ^{71}{\rm Ge}\xrightarrow{\text{EC}}\ ^{71}{\rm Ga}$ is suitable to determine the thermal neutron fluence due to its accompanying KX-ray emissions. The $^{71}$Ge has a half-life of 11.43 days and decays solely by electron capture. The EC process produced atomic K-shell vacancies, and results in emission of Ga characteristic low-energy X-rays of 9.2 keV and 10.3 keV. TABLE~\ref{tab:table0905_2} shows the X-rays radiation of this marked decay channel, and the data were taken from \cite{nucleardata}. 
Owing to that $^{71}$Ge is the most abundant nuclide produced in NTD processes and has the relatively long half-life, it makes people have comfortable time schedule to measure the X ray spectra. In this report, $^{71}$Ge is chosen as monitor’s nuclide and the cascading $K_\alpha$, $K_\beta$ X-rays of Ga (Table ~\ref{tab:table0905_1}) are detected to tag the EC-decays of $^{71}$Ge with the suitable detectors. 

\begin{table}[h!]
\caption{\label{tab:table0905_2}Information of adjoint Ga X-rays of $^{71}$Ge electron capture. }
\begin{tabular*}{\tblwidth}{@{}LLL@{}}
\toprule
&\multicolumn{1}{c}{\textrm{Energy}} &\multicolumn{1}{c}{\textrm{Intensity\footnote{Absolute intensity per 100 parent nuclei $^{71}$Ge decays.}}}\\
&\multicolumn{1}{c}{(\textrm{keV})} &\multicolumn{1}{c}{(\%)}\\
\midrule
K-L$_2$ &9.225 &13.43(13)\\
K-L$_3$ &9.253 &26.10(21)\\
K-M$_{2,3,4}$ &10.260-10.351 &5.73(7)\\
K-MN &10.260-10.365 &5.76(7)\\
\bottomrule
\end{tabular*}
\end{table}
According to the dynamic theory of nuclide transmutation and radionuclide decay, the relationship between neutron flux ($\bar{\phi}$) and $^{71}$Ge radioactivity $A(t_\text{cool})$ can be written as:
\begin{equation}
A(t_\text{cool})=A(t_r)e^{-\lambda t_\text{cool}}=\sigma_0\bar{\phi} N_0 (1-e^{-\lambda t_r})e^{-\lambda t_\text{cool}},
\label{eq:0905_1}
\end{equation}
where $t_\text{cool}$ is the cooling time after irradiation, $t_r$ is the duration time of the irradiation, $\lambda$ is the decay constant of $^{71}$Ge, $\sigma_0$ is the neutron capture cross-section of $^{70}$Ge measured at 2200~m/s, $\bar{\phi}$ is the average thermal neutron flux under the Westcott convention\cite{osti_4118414} and $N_0$ is the initial number of the $^{70}$Ge nuclides in the HP-Ge sample. The neutron fluence $\Phi$ (product of neutron flux $\bar{\phi}$ and $t_r$) can be derived from the measured activity $A(t_\text{cool})$ of $^{71}$Ge. By measuring the counting rate of Ga KX-rays and combining the detection efficiency acquired using GEANT4 simulation, the radioactivity $A(t_\text{cool})$ of $^{71}$Ge can be determined and the average neutron flux can be acquired, the neutron fluence $\bar{\phi}\cdot t_r$ can be calculated. Considering that the energy of detected X-rays is approximately 10~keV, an Micro-Megas Detector (MMD) and a Silicon Drift Detector (SDD) can meet the requirements. 

\section{Experiments, simulations and results}\label{sec::0905_3}
We prepared 10N HP-Ge wafer of 1-mm thickness purchased from UMICORE\cite{umicore.org} and cut it by a wafer dicing machine. 20 pieces of 10~mm$\times$10~mm$\times$1~mm HP-Ge in total were packed into a high-purity thin quartz box. The 20 boxes are then divided into seven groups: 6 groups of 3 boxes, 1 group of 2 boxes. Each group is sealed into a high-purity Aluminum (Al) can and is placed in a unique irradiation site. Seven groups with 20 pieces of HP-Ge are respectively located at seven sites in the irradiation channel of China Advanced Research Reactor(CARR). By controlling the power of reactor and the irradiation time, so as seven group of Ge-samples have their preset neutron fluences in the range of about $3\sim9 \times 10^{18}$ n$\cdot$ cm$^{-2}$. The share of thermal neutron of the irradiation site in CARR is greater than 99 percent by Cadmium-ratio method. After the irradiation, the NTD-Ge samples were kept for more than 150 days to deplete the short-term radioactive isotopes.  All the samples were wrapped in polyethylene film (thickness: 10 $\mu$m) before measuring their activities to prevent from cross-contamination.

\subsection{MMD measurement}
\begin{figure}[!htbp]
\begin{center}
\subfigure[]{
\includegraphics[height=3.5cm]{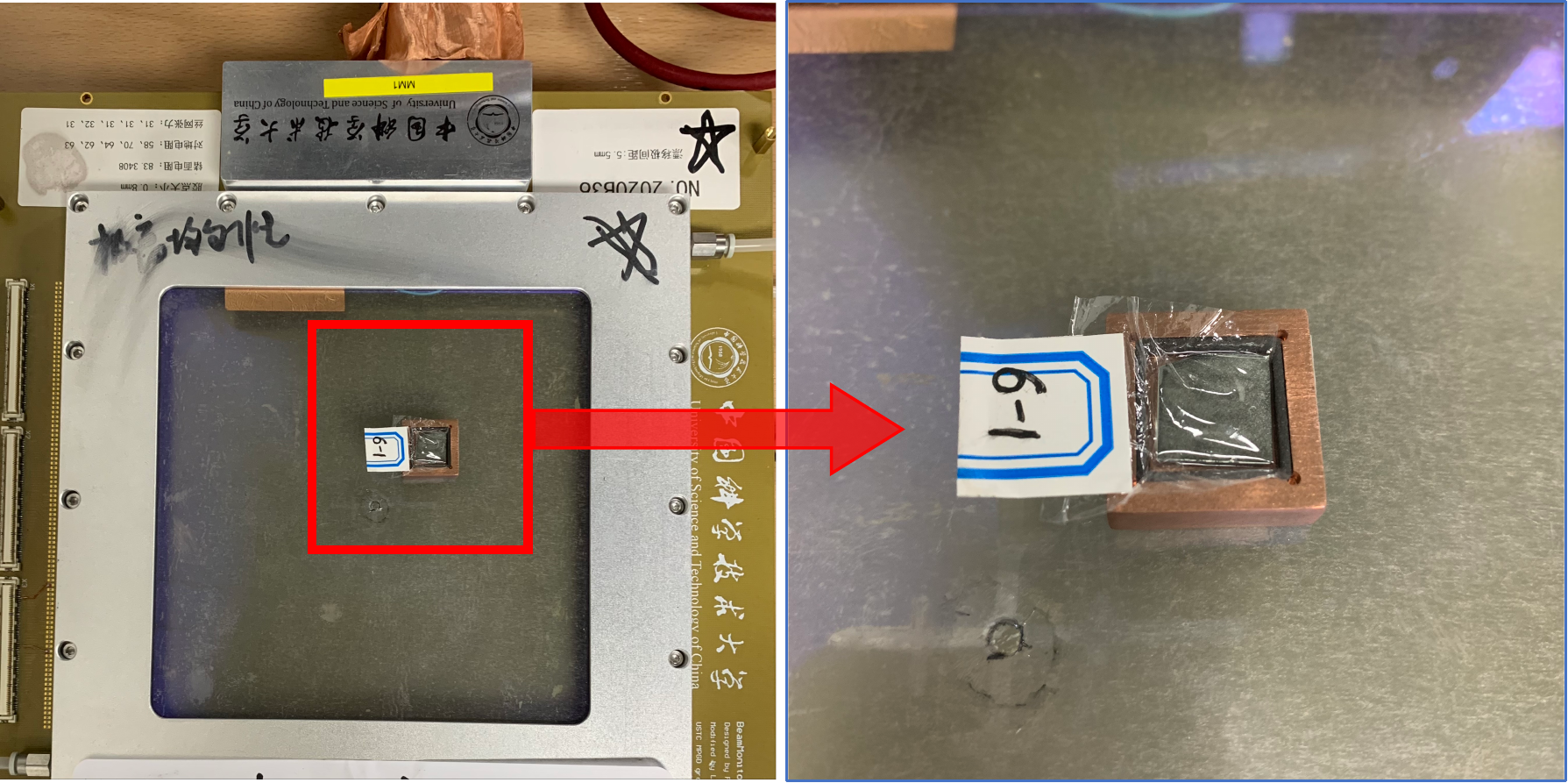}}
\subfigure[]{
\includegraphics[height=3.5cm]{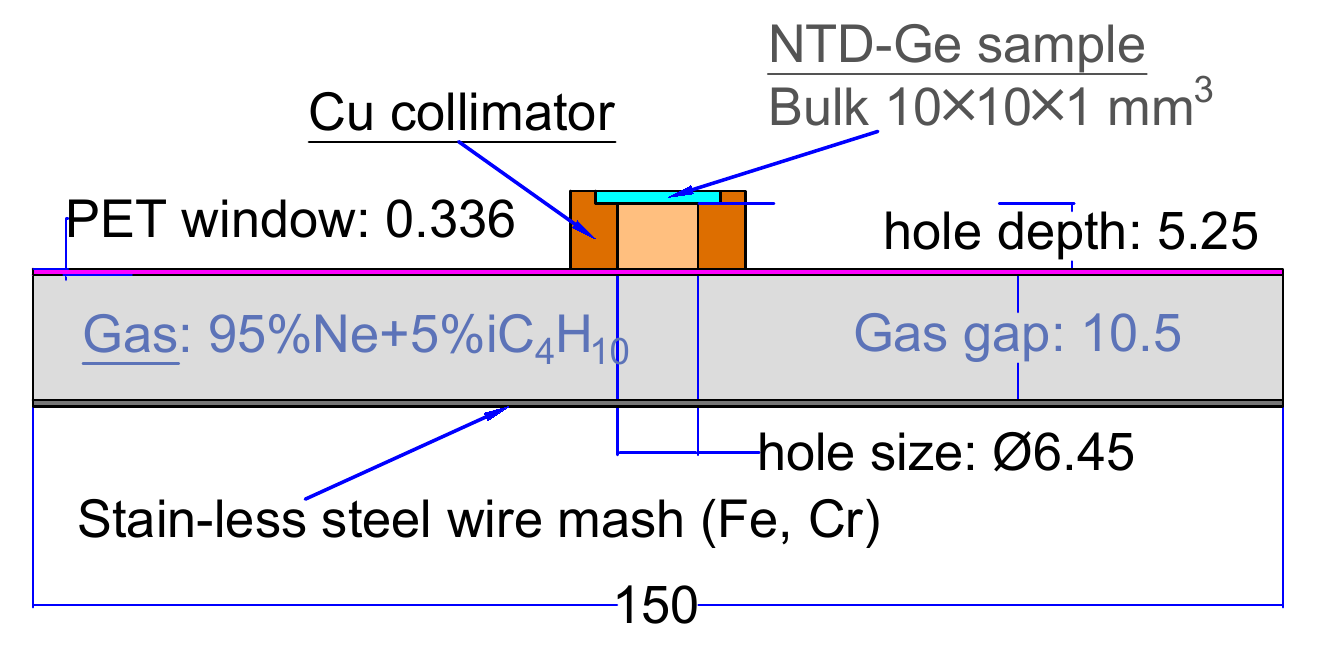}}
\caption{\label{fig:0905_3} MMD experimental setup used to measure the Ga KX-ray spectra of NTD-Ge. (a) Photograph of the MMD physical setup. (b) MMD schematic diagram, dimensions are given in mm.}
\end{center}
\end{figure}

\begin{figure}[!htbp]
\begin{center}
\includegraphics[width=8.5cm]{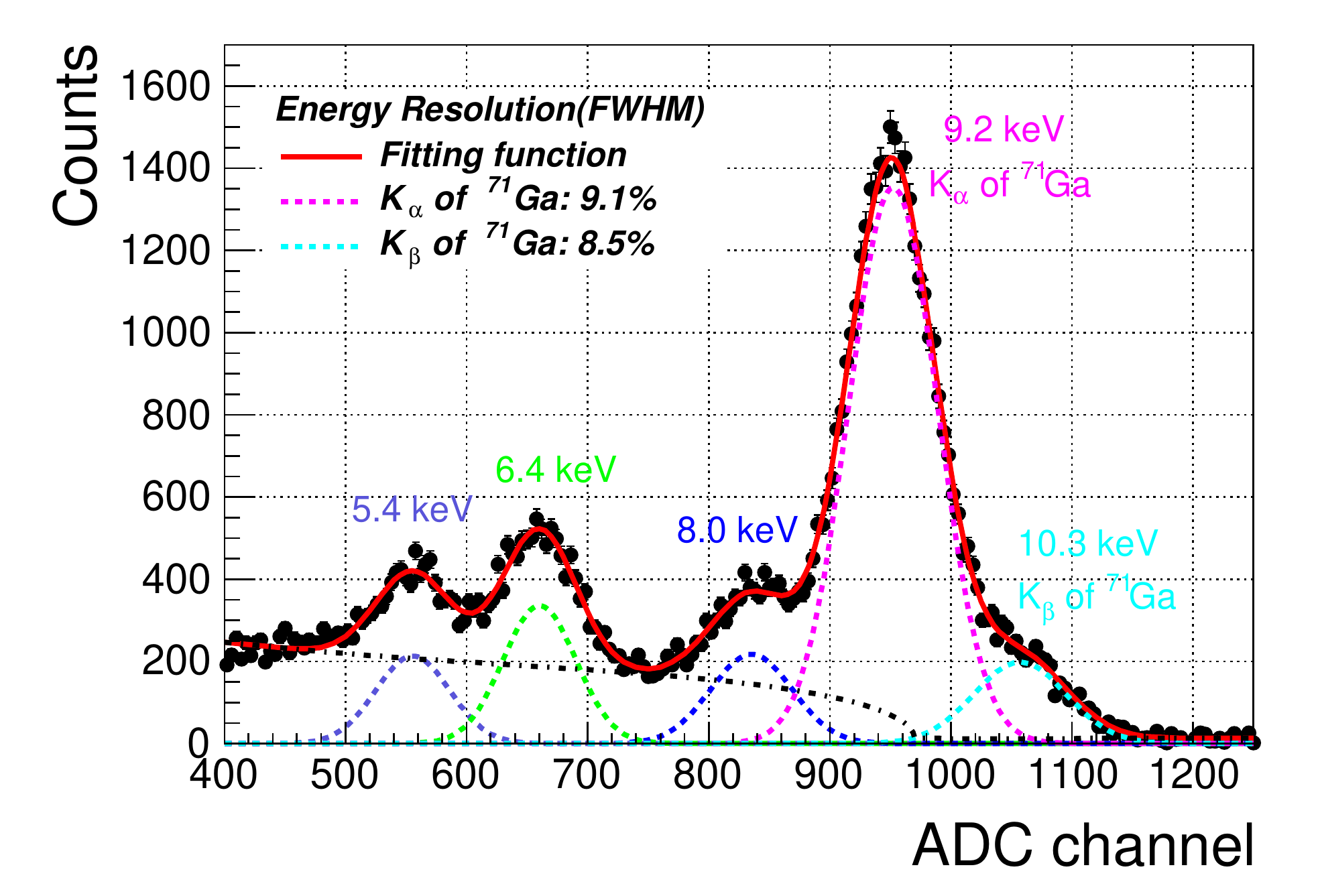}
\caption{\label{fig:0905_4} An energy spectrum of an NTD-Ge sample measured by MMD. Besides for the KX-rays from NTD-Ge with 9.2~keV and 10.3~keV, there are still three energy peaks that come from the atomic de-excitation of Cu in copper collimator (8.0~keV) and Cr (5.4~keV), Fe (6.4~keV) in stainless-steel wire mesh. The Argus background platform is from multiple scattering of electrons.}
\end{center}
\end{figure}

The MMD we used is manufactured by a thermal bonding method with high detector performance\cite{feng2021thermal,FENG2022166595}. It has an active area of 15$\times$15 cm$^2$  and 1.05-cm thickness gas chamber as the drift region for primary ionization. In this measurement, the mixture gas with 95\% Neon(Ne) + 5\% iC$_4$H$_{10}$ is used to obtain a better energy resolution with Full Width at Half Maximum (FWHM) of 11.6\% at 5.9~keV. 
In order to have well defined acceptance for measuring the X-rays, a Copper (Cu) collimator is used during the test. At the same time, the collimator is placed at the center of the sensitive area of the MMD to avoid the edge effect of the detector. FIG.~\ref{fig:0905_3} displays the MMD experimental setup used to measure the Ga KX-ray spectra of the NTD-Ge sample. In the schematic diagram, only the parts that determine the detection efficiency and X-ray eneygy spectra are shown. (see Ref.\cite{feng2021thermal,FENG2022166595}.)
The MMD will record the photoelectric absorptions in the gas of the incident X-rays as monoenergetic peaks, which is only a fraction of the X-rays cascading from $^{71}$Ge dacays. Most X-rays have no interaction with the gas because they deposit all the energy by photoelectric effect in other components such as the NTD-Ge itself, Cu collimator, PET window and the stainless-steel mesh. The energy spectrum of an NTD-Ge sample measured by MMD is shown in FIG.~\ref{fig:0905_4}, which is fitted by a sum of five Gaussian functions (to describe kX-rays) and an Argus function (to describe multiple scattering background) simultaneously. The MMD is calibrated using $^{55}$Fe with X-ray of 5.9~keV, and give a response about 100 ADC channel per keV.
The $K_\alpha$ (9.2~keV) and $K_\beta$ (10.3~keV) peaks of Ga are the energy peaks of interest. In addition, there are three prominent energy peaks at 5.4~keV, 6.4~keV, and 8.0~keV, which are introduced due to the material atomic de-excitation of Chromium (Cr), Iron (Fe) in stainless-steel wire mesh and Copper (Cu) in collimator. When the incident X-rays produce atomic shell vacancies of these atoms through the photoelectric effect in matter, the further de-excitations emmit these characteristic X-rays. However, these three energy peaks have no competition with the Ga X-ray energy peaks, we are only concern about the events from the direct interaction possibility of Ga X-rays with gas, which were recorded as 9.2~keV and 10.3~keV. According to the fit results, we get the measured count of each X-ray which is the integral area of each gaussian peak. 

The GEANT4 Monte Carlo simulation toolkit \cite{agostinelli2003geant4} is used to calculate the MMD detection efficiency. The efficiency we need to know is the number of MMD maseured X-rays per $^{71}$Ge dacay in the NTD-Ge sample. The geometric parameters of the MMD shown in FIG.~\ref{fig:0905_3}(b) and the atomic data of all involved materials are implemented in the simulation. A certain number ($N_0$) of $^{71}$Ge nuclides were randomly and homogeneously generated with zero kinetic energy in the NTD-Ge bulk (10$\times$10$\times$1 mm$^3$) by using the G4GeneralParticleSource class. 
These conditions in simulations are regarded as consistent with the experiments, since the NTD technique can produce a very uniform doping profile. And by using the Cu collimator, the little deviation of the size and position of bulk NTD-Ge in simulation and experiments make no difference on the final result of detection efficiency.
The decay processes of $^{71}$Ge were then simulated by the G4RadioactiveDecayPhysics class, this class will simulate the whole process associated with the $^{71}$Ge decay according to the Evaluated Nuclear Structure Data File (ENSDF)\cite{tuli1996evaluated}. And finally generate the X-ray emissions with the same intensities as shown in TABLE~\ref{tab:table0905_2}. The emissions of cascading X-rays are isotropic. The atomic de-excitation is activated by switching on the Particle Induced X-ray Emission (PIXE), Auger and Fluorescence emissions. 
GEANT4 version with 4-10.2.1 and CLHEP libraries version with 2.3.1.0 are used. In addition to the standard electromagnetic process, there are two specific low energy electromagnetic models available, Livermore and Penelope\cite{lowenergyModel}, which are optimized to describe low energy particles. During the simulations, Livermore, Penelope and EMStandard\_option4 packages are used, and all these models give consistent results for the detection efficiency calculation. We gave up the X-rays that were absorbed by the Cu collimator or the sample itself, and only traced the survival X-rays which have passed through the PET window. If X-rays interact with working gas and produce photo-electron, we counted the numbers  $N_{pe}$  of photoelectron whose energies have completely transferred to produce the primary electron-ion pairs in the drift region of MMD. By perfectly constructing the measurement system, we take all the factors into account together and get the final efficiency: from the activity of the $^{71}$Ge in the NTD-Ge sample to the counting rate of the detector, $\eta_{\text{MMD}}$=$\frac{N_{pe}}{N_0}$. Limited by the energy resolution of MMD as shown in FIG~\ref{fig:0905_4}, we only calculated the detection efficiency for 9.2~keV X-rays to estimate the activities of $^{71}$Ge: $\eta_{\text{MMD}}(9.2~\text{keV})$=(3.584$\pm$0.134)$\times 10^{-6}$. 

\subsection{SDD measurement}
\begin{figure}[!htbp]
\begin{center}
\includegraphics[width=8.5cm]{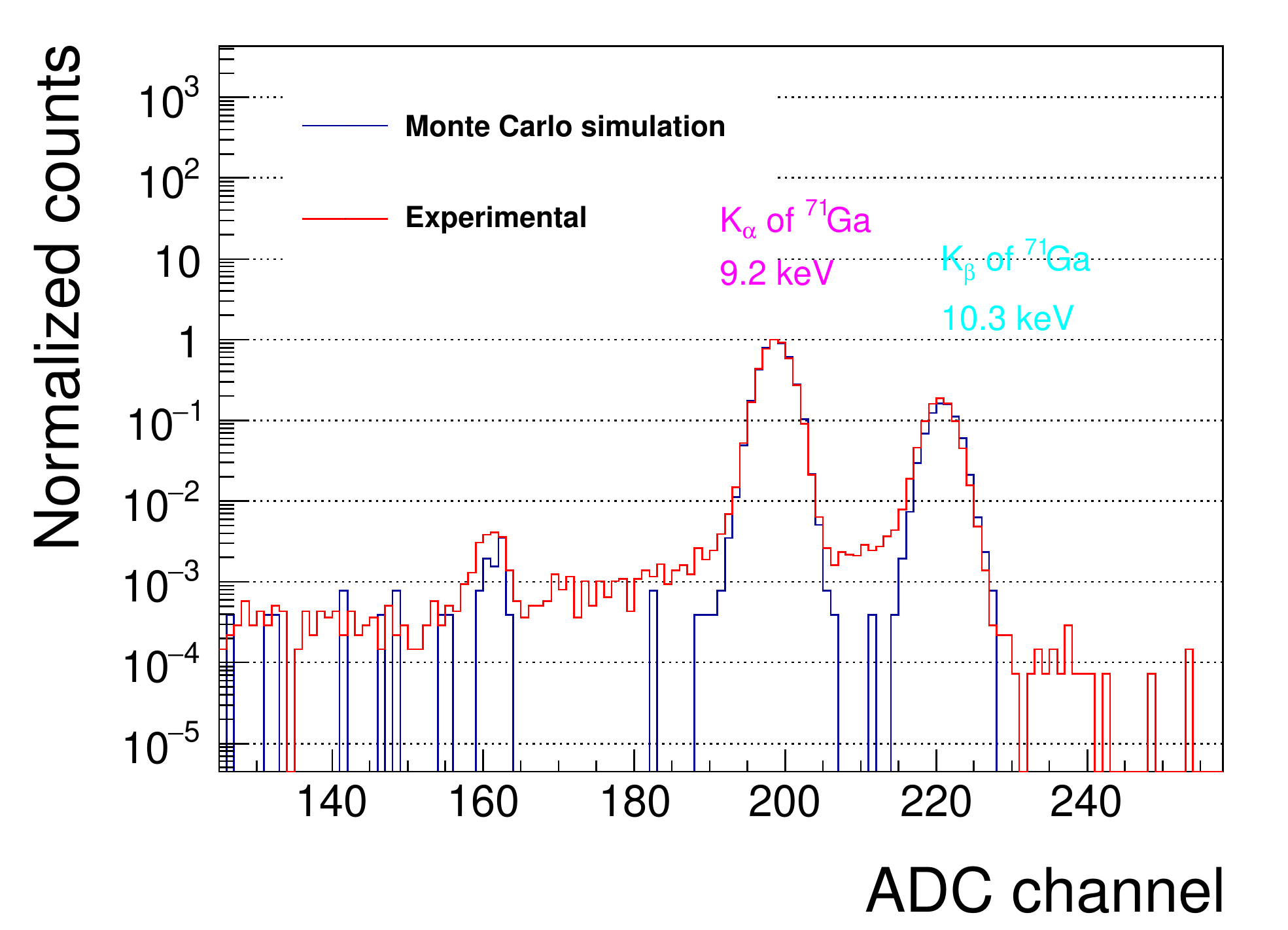}
\caption{\label{fig:0905_6} Energy spectrum of an NTD-Ge sample measured by SDD. The red line is experimentally measured and the blue line is GEANT4 simulated. The simulated spectra have been convoluted by a Gaussian function to account for the limited detector resolution, i.e. 150~eV in FWHM at 5.9~keV. This graph shows that the spectra of the simulation and experiments are in a good agreement.}
\end{center}
\end{figure}

To improve the detection accuracy and perform the cross check, a semiconductor detector, SDD, is used in this measurement as well. This SDD is a integrated, commercial detector produced by Amptek company\cite{timmurphy.org}, which can offer a magnificent energy resolution with FWHM $\sim$150~eV at 5.9~keV. Because there is a internal collimator inside the detector probe, no extra Cu collimator is used during the test. The energy spectrum of an NTD-Ge sample is shown in FIG.~\ref{fig:0905_6}. Due to the high energy resolution of SDD, there are two separate energy peaks from Ga KX-rays with 9.2~keV and 10.3~keV clearly, and the results show that the relative intensities of $K_\alpha$ and $K_\beta$ by simulation and experiment are in good agreement. The GEANT4 simulations are also used to give the SDD detection efficiency. The NTD-Ge samples were placed on the surface of the Be window during the experiment. And the SDD detection efficiency of GEANT4 simulated is $\eta_{\text{SDD}}$=(5.592$\pm$0.116)$\times 10^{-4}$. This value is convincing after verifying that the efficiencies of GEANT4 simulations, factory report\cite{sddeff.org} and the experimental measurements (using a Monoenergetic X-Ray Calibration Device)\cite{2019-T08} are all consistent with each other. 
\subsection{The results}
\begin{figure}[!htbp]
\begin{center}
\includegraphics[width=8.5cm]{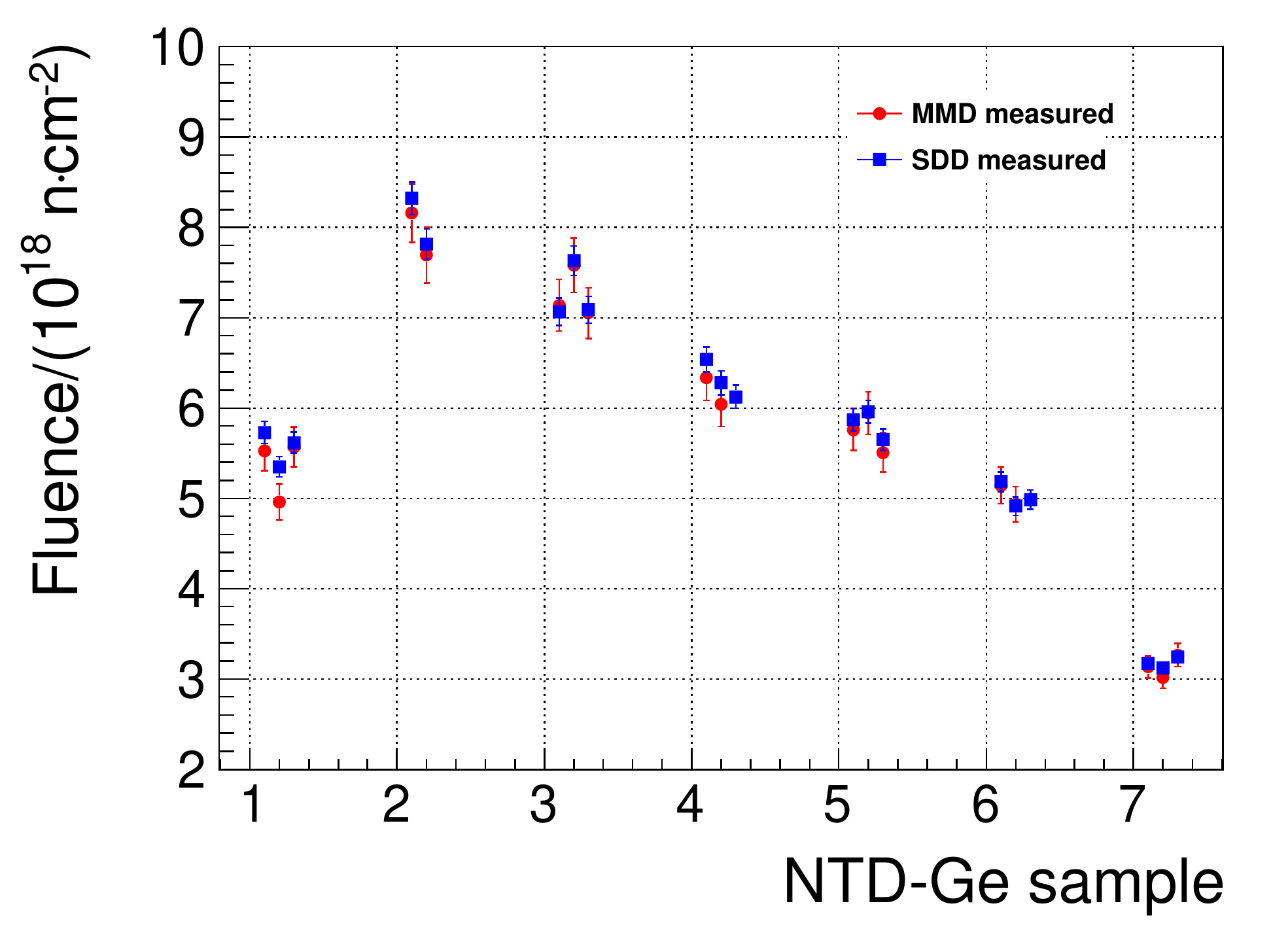}
\caption{\label{fig:0905_7}Neutron fluence of all the NTD-Ge samples, the results of MMD and SDD measurements coincide with each other. The error bars do not include the contribution of input parameters which are same for both detectors.}
\end{center}
\end{figure}

Based on the measured counting rates of K-shell X-rays using MMD and SDD, and the GEANT4 simulated efficiencies, we derive the $_{32}^{71}$Ge-radioactivity of each NTD-Ge sample. According to the Eq.~\ref{eq:0905_1}, the average neutron fluence can be acquired. FIG.~\ref{fig:0905_7} shows the neutron fluence for each NTD-Ge using data from MDD and SDD, which have good agreement with each other. This consistency also indicate that we build correct geometric and physical structures for both MMD and SDD.
The horizontal axis represents seven irradiation groups numbered 1-7. There are 18 fluence points from MMD measurements and 20 fluence points from SDD measurements on the plot. The sources of systematic uncertainties are considered in the neutron fluence calculation. The uncertainties from the peak fit for $^{71}$Ga kX-rays are studied by alternative fits with different fit ranges, background parameters, and the energy resolution changes. Due to the energy resolution and interaction process, the uncertainties in estimating the counting rate of MMD and SDD are 1.321\% and 0.428\% respectively. In terms of the detection efficiencies of MMD and SDD, errors in the detector structure such as the size and positions of all the components will contribute to the uncertainties of the simulated results. All these uncertainties are combined in the simulations, and the obtained uncertainties for MMD and SDD are 3.755\% and 2.078\%, respectively. These above two terms indicate that the uncertainties in $^{71}$Ge activity are 3.98\% and 2.12\% for MMD and SDD measurements, respectively. The uncertainties contributed by the input parameters, such as the half-life time of $^{71}$Ge, the $^{70}$Ge isotope natural abundance and the dominantly effective thermal neutron capture cross-section of $^{70}$Ge are also taken into consideration. Together, these parameters contribute nearly 5.9\% uncertainties to the neutron fluence results of MMD and SDD.
The individual uncertainties are summarized in TABLE ~\ref{tab:table0905_3}. 
\begin{table}[!htbp]
\caption{\label{tab:table0905_3}Uncertainty budget in the measurement of thermal neutron fluence.}
\begin{tabular*}{\tblwidth}{@{}LLL@{}}
\toprule
\multirow{2}{*}{Uncertainty item}&\multicolumn{2}{c}{Value in \%}\\ \cline{2-3}
&\multicolumn{1}{c}{MMD}&\multicolumn{1}{c}{SDD} \\
\midrule
Peaks fit&1.321&0.428\\
Efficiency&3.755&2.078\\
Input parameters&5.957&5.905\\\hline
\bottomrule
\end{tabular*}
\end{table}

\section{Conclusions and prospects}\label{sec::0905_4}
We have demonstrate a method that using the activation Ge itself to determine the thermal neutron fluence of NTD-Ge. The commonly used activation monitors (Co and Au) can not afford a high thermal neutron fluence up to 10$^{18}$~n$\cdot$cm$^{-2}$ due to their relative large neutron capture cross-section and long-lived radionuclides. The reaction process $^{70}{\rm Ge}(n, \gamma)\rightarrow\ ^{71}{\rm Ge}\xrightarrow{\text{EC}}\ ^{71}{\rm Ga}$ has suitable
nuclear parameters and emits X-rays for experimental measurements. A Micro-Megas Detector with high detector performance and a Silicon Drift Detector with high energy resolution are used to measure the spectra of $^{71}$Ga KX-rays (adjoint X-ray of the K-shell electron capture process of radioactive nuclide $^{71}$Ge) with energies of 9.2~keV and 10.3~keV. Both the detection efficiencies of MMD and SDD are obtained by GEANT4 simulations. The thermal neutron fluence results with proper accuracy of NTD-Ge samples given by these two detectors are in good agreement. The overall accuracy of the germanium activation monitor is limited by the accuracy of the neutron capture cross-section of $^{70}$Ge, which can be improved if it is studied. The correction for epithermal activation and sample self-shield are not considered in our experiments. Self-monitor method (presented above) could offer the real thermal neutron fluence of an irradiation sample, but extra monitor method could only offer the approximately thermal neutron fluence of an irradiated sample. Looking for a proper nuclear chain of a irradiated sample is a key step for using self-monitor method successfully. 
\section{Aknowledgement}
The authors thank the CARR for its strong support on HP-Ge irradiation. We thank Guo Siming from China National Institute of Metrology for his help on calibration of SDD detection efficiency. The HP-Ge wafer dicing and surface treatment were performed at the University of Science and Technology of China (USTC) Center for Micro and Nanoscale Research and Fabrication, and we thank Yu Wei and Haitao Liu for their kind help. Special thanks to Prof.Zizong Xu, Prof.Qing Lin from USTC for their kindly support and helpful discussion.
This work was supported bythe National Natural Science Foundation of China(1200522511721505, 1162552361571411) the Fundamental Research Funds for the Central Universities WK2030000024), and the State Key Laboratoryof Particle Detection and Electronics(SKLPDE-ZZ-202012).




\bibliographystyle{elsarticle-num} 

\bibliography{myfile}

\begin{thebibliography}{10}
\expandafter\ifx\csname url\endcsname\relax
  \def\url#1{\texttt{#1}}\fi
\expandafter\ifx\csname urlprefix\endcsname\relax\def\urlprefix{URL }\fi
\expandafter\ifx\csname href\endcsname\relax
  \def\href#1#2{#2} \def\path#1{#1}\fi

\bibitem{barabash2000neutron}
V.~Barabash, G.~Federici, M.~R{\"o}dig, L.~Snead, C.~Wu, Neutron irradiation
  effects on plasma facing materials, Journal of Nuclear Materials 283 (2000)
  138--146.

\bibitem{hasegawa2013neutron}
A.~Hasegawa, M.~Fukuda, T.~Tanno, S.~Nogami, Neutron irradiation behavior of
  tungsten, Materials transactions (2013) MG201208.

\bibitem{bortolussi2007thermal}
S.~Bortolussi, S.~Altieri, Thermal neutron irradiation field design for boron
  neutron capture therapy of human explanted liver, Medical physics 34~(12)
  (2007) 4700--4705.

\bibitem{janus1976application}
H.~M. Janus, O.~Malmros, Application of thermal neutron irradiation for large
  scale production of homogeneous phosphorus doping of floatzone silicon, IEEE
  Transactions on Electron Devices 23~(8) (1976) 797--802.

\bibitem{shlimak1999neutron}
I.~Shlimak, Neutron transmutation doping in semiconductors: science and
  applications, Physics of the Solid State 41~(5) (1999) 716--719.

\bibitem{larrabee2013neutron}
R.~D. Larrabee, Neutron transmutation doping of semiconductor materials,
  Springer Science \& Business Media, 2013.

\bibitem{kohler1971determination}
W.~K{\"o}hler, R.~Vaninbroukx, The determination of the thermal neutron fluence
  by cobalt activation monitors, The International Journal of Applied Radiation
  and Isotopes 22~(9) (1971) 529--541.

\bibitem{simonits1976zirconium}
A.~Simonits, F.~De~Corte, J.~Hoste, Zirconium as a multi-isotopic flux ratio
  monitor and a single comparator in reactor-neutron activation analysis,
  Journal of Radioanalytical and Nuclear Chemistry 31~(2) (1976) 467--486.

\bibitem{de1991calibration}
F.~De~Corte, A.~De~Wispelaere, R.~Jonckheere, et~al., Calibration of the
  fission-track dating method: is cu useful as an absolute thermal neutron
  fluence monitor?, Chemical Geology: Isotope Geoscience section 86~(3) (1991)
  187--194.

\bibitem{201889}
\href{https://www.sciencedirect.com/science/article/pii/B9780444637697000154}{Recommended
  thermal cross sections, resonance properties, and resonance parameters for z
  = 1–60}, in: S.~Mughabghab (Ed.), Atlas of Neutron Resonances (Sixth
  Edition), sixth edition Edition, Elsevier, Amsterdam, 2018, pp. 89--822.
\newblock \href
  {https://doi.org/https://doi.org/10.1016/B978-0-44-463769-7.00015-4}
  {\path{doi:https://doi.org/10.1016/B978-0-44-463769-7.00015-4}}.
\newline\urlprefix\url{https://www.sciencedirect.com/science/article/pii/B9780444637697000154}

\bibitem{isotopicdata}
Germanium isotopic abundance,
  \url{https://physics.nist.gov/cgi-bin/Compositions/stand_alone.pl?ele=Ge},
  accessed August, 2021.

\bibitem{nucleardata}
Decay radiation of ge-71 ec decay from iaea database,
  \url{https://www-nds.iaea.org/relnsd/vcharthtml/VChartHTML.html}, accessed
  Apirl, 2022.

\bibitem{osti_4118414}
C.~H. Westcott, \href{https://www.osti.gov/biblio/4118414}{Effective cross
  section values for well-moderated thermal reactor spectra. (3rd edition
  corrected)}.
\newline\urlprefix\url{https://www.osti.gov/biblio/4118414}

\bibitem{umicore.org}
Hp-ge products from umicore company,
  \url{https://eom.umicore.com/en/germanium-solutions/products/high-purity-germanium-crystals/
  }, accessed Apirl, 2022.

\bibitem{feng2021thermal}
J.~Feng, Z.~Zhang, J.~Liu, B.~Qi, A.~Wang, M.~Shao, Y.~Zhou, A thermal bonding
  method for manufacturing micromegas detectors, Nuclear Instruments and
  Methods in Physics Research Section A: Accelerators, Spectrometers, Detectors
  and Associated Equipment 989 (2021) 164958.

\bibitem{FENG2022166595}
J.~Feng, Z.~Zhang, J.~Liu, M.~Shao, Y.~Zhou,
  \href{https://www.sciencedirect.com/science/article/pii/S0168900222001711}{A
  novel resistive anode using a germanium film for micromegas detectors},
  Nuclear Instruments and Methods in Physics Research Section A: Accelerators,
  Spectrometers, Detectors and Associated Equipment 1031 (2022) 166595.
\newblock \href {https://doi.org/https://doi.org/10.1016/j.nima.2022.166595}
  {\path{doi:https://doi.org/10.1016/j.nima.2022.166595}}.
\newline\urlprefix\url{https://www.sciencedirect.com/science/article/pii/S0168900222001711}

\bibitem{agostinelli2003geant4}
S.~Agostinelli, J.~Allison, K.~a. Amako, J.~Apostolakis, H.~Araujo, P.~Arce,
  M.~Asai, D.~Axen, S.~Banerjee, G.~Barrand, et~al., Geant4—a simulation
  toolkit, Nuclear instruments and methods in physics research section A:
  Accelerators, Spectrometers, Detectors and Associated Equipment 506~(3)
  (2003) 250--303.

\bibitem{tuli1996evaluated}
J.~K. Tuli, Evaluated nuclear structure data file, Nuclear Instruments and
  Methods in Physics Research Section A: Accelerators, Spectrometers, Detectors
  and Associated Equipment 369~(2-3) (1996) 506--510.

\bibitem{lowenergyModel}
Geant4 toolkit official website, \url{https://geant4.web.cern.ch/node/148},
  accessed August, 2021.

\bibitem{timmurphy.org}
\url{https://www.amptek.com/internal-products/obsolete-products/sdd-x-ray-detectors-for-xrf/x-123sdd-complete-x-ray-spectrometer-with-silicon-drift-detector-sdd},
  accessed August, 2022.

\bibitem{sddeff.org}
Product details and structural information of x-123 sdd from amptek company,
  \url{https://www.amptek.com/-/media/ametekamptek/documents/resources/efficiency.zip?la=en&revision=b1f6f184-3765-43db-b12b-9f8b32a40ae2},
  accessed August, 2022.

\bibitem{2019-T08}
Z.~J. GUO~Siming, WU~Jinjie, Introduction of monoenergetic x-ray calibration
  device and test items of nim, National Institute of Metrology, Beijing
  100029, China 65~(3) (2021) 3--8.

\end{thebibliography}
\bio{}
\endbio

\end{document}